\documentclass[twocolumn]{article}          
\usepackage{graphicx}
\usepackage{amsmath}
\begin{document}

\title{{\bf Comparing Machine Learning Algorithms with or without Feature Extraction for DNA Classification}
}


\author{Xiangxie Zhang$^1$         \and
        Ben Beinke$^1$             \and
        Berlian Al Kindhi$^2$ \and
        Marco Wiering$^1$ \\
       $^1$ Dept. of Artificial Intelligence, University of Groningen, The Netherlands \\
       $^2$ Dept. of Electrical Automation Engineering, Institut Teknologi Sepuluh Nopember, Indonesia
}



\date{}

\maketitle

\begin{abstract}
The classification of DNA sequences is a key research area in bioinformatics as it enables researchers to conduct genomic analysis and detect possible diseases. In this paper, three state-of-the-art algorithms, namely Convolutional Neural Networks, Deep Neural Networks, and N-gram Probabilistic Models, are used for the task of DNA classification. Furthermore, we introduce a novel feature extraction method based on the Levenshtein distance and randomly generated DNA sub-sequences to compute information-rich features from the DNA sequences.
We also use an existing feature extraction method based on 3-grams to represent amino acids and combine both feature extraction methods with a multitude of machine learning algorithms. Four different data sets, each concerning viral diseases such as Covid-19, AIDS, Influenza, and Hepatitis C, are used for evaluating the different 
approaches. The results of the experiments show that all methods obtain 
high accuracies on the different DNA datasets. Furthermore, the domain-specific 3-gram feature extraction method leads in general
to the best results in the experiments, while the newly proposed
technique outperforms all other methods on the smallest Covid-19 dataset.


\end{abstract}

\section{Introduction}
\label{intro}
The first successful isolation of DNA by Friedrich Miescher in 1869 was a groundbreaking step in biology as it laid the groundwork of understanding the blueprints of all organic life. DNA, which is short for deoxyribonucleic acid, is a hereditary material that can be found in the cells of all humans and other living organisms. It carries the necessary information which decides the biological traits of our bodies and works as a genetic blueprint for an evolving organism. An isolated DNA sequence can be represented by a character string, which consists of only A, C, G, or T. This format is named FASTA. The analysis of DNA is crucial, as it allows doctors to diagnose diseases, helps in analyzing the spread of new infections, and it can also be used to solve crimes or conduct paternity tests. Therefore, DNA analysis has become a vital interest in computational biology \cite{gelfand1995prediction}.

In traditional biology, primers are essential tools for DNA analysis. Primers are short single-stranded nucleotide sequences important for the initiation phase of DNA synthesis of all living organisms. In molecular biology, synthetic primers are utilized for different purposes such as the detection of viruses \cite{bukh1992importance}, bacteria \cite{dorn2015specific} or parasites \cite{pacheco2018primers}. Primers, which are often present in human DNA sequences that are infected by a specific type of virus, are utilized for these purposes. With the help of the Polymerization Chain Reaction (PCR) method, the DNA fragment of the existing virus is amplified significantly, and researchers are able to detect the virus. 

Primers are also utilized in various DNA classification problems \cite{muller1997classification,porter2012factors} and bioinformatics \cite{barnes2007bioinformatics}. For this study, it is important to note that they can be considered as comparison patterns that can be searched for to diagnose diseases. By calculating the edit distances between an isolated DNA sequence and the primers of a specific virus, the level of the virus being expressed in the human DNA sequence can be obtained, which can then be used to build up the feature vectors. Machine learning algorithms can be trained on the feature vectors of the DNA sequences.  The resulting model can be used to detect viruses and diagnose viral diseases.

{\bf Contributions.} Synthesizing primers for a particular virus is often difficult and expensive. This paper proposes an alternative method that uses randomly generated DNA sequences to replace the primers. The advantage is that the analysis and processing necessary for finding the primer patterns can be ignored. In the experiments, the performances of feature extraction using primers and random DNA sequences will be compared to several other machine learning approaches. Another feature extraction method, which will be referred to as the 3-gram method throughout this paper, is also developed. Additionally, other state-of-the-art algorithms, namely convolutional neural networks \cite{lecun1989generalization,lecun1998gradient} (CNNs) and deep neural networks (DNNs), which extract the features directly from sample DNA sequences, are evaluated. They are compared with the two feature extraction methods combined with machine learning algorithms such as Adaboost \cite{freund1995desicion}, support vector machines (SVMs) \cite{boser1992training,vapnik:1995}, and others. An additional algorithm, the N-gram probabilistic model \cite{tomovic2006n,chen2014private}, which is often used in natural language processing, is also implemented and compared to the other machine learning approaches.

To provide accurate and convincing final results, we conducted each of the experiments and tested all methods on four different data sets. One data set is concerned with the detection of the Hepatitis C virus in human DNA, one with the classification of influenza virus and coronavirus, another with the classification of HIV into HIV type 1 and HIV type 2, and the last with a classification problem based on human DNA samples infected with SARS-Cov-2. A detailed description of the data sets can be found in the subsequent section. 

{\bf Paper outline.} This paper is organized as follows. In Section 2, each of the four data sets is described, and the decisions that motivated the choices of data are explained. Section 3 explains the used feature extraction methods and machine learning algorithms. This is followed in Section 4 by the description of the experimental setup. The results for each experiment are presented in Section 5, while Section 6 discusses the results and concludes the paper.

\section{Datasets}

The DNA sequence classification methods are tested on four different data sets of various sizes. They include different types of viruses, and different datasets were used for different aims. One data set that is commonly used and might be considered the standard for DNA analysis by some, the Molecular Biology (Splice-junction Gene Sequences) Data Set, was not used. This decision was made because of the length of the samples this data set contains. While the samples of the tested data sets contained up to almost 30,000 characters, the samples of the splice data set are only sequences of 61 characters. This reason, in addition to the age of the data, the data set was created in 1992, led to the decision of using newer data sets with samples more resembling data encountered in actual applications. 

\subsection{Hepatitis C virus dataset}\label{sec:dataset}

The hepatitis C virus (HCV) is a single-stranded RNA virus that can infect RNA sequences in the human body. RNA is the messenger that contributes to the formation of DNA. Therefore, if the RNA is infected, the DNA is also modified. Unlike the hepatitis B virus (HBV), an effective vaccine against HCV has not yet been developed \cite{grady2015hepatitis}. HCV can cause severe diseases like hepatitis C and liver cancer. Thus it is vital to detect potential infections with HCV as early as possible. This HCV dataset was obtained from the World Gene bank and consists of 500 HCV positive DNA sequences and 500 HCV negative DNA sequences. The length of the DNA sequences in this data set varies widely, with the longest sequences being 12,722 characters long and the shortest only 73 characters long. Most sequences, however, fall in the range from 9,000 to 12,000.
\subsection{Influenza \& corona virus dataset}
The coronavirus is an RNA virus that can infect humans' respiratory tract and cause many different diseases. Potential diseases could be mild like the common cold, but it could also be lethal like SARS, MERS, or Covid-19. On the other hand, the influenza virus is responsible for seasonal flu and has caused many epidemics in history; for example, the Spanish influenza in 1918 and the outbreak of H1N1 in 2009. Influenza viruses and coronaviruses may cause similar symptoms to patients. However, different measures might need to be taken in order to support patients in their recoveries, depending on the type of virus they are infected with. Therefore it is crucial to know which kind of infection a patient has before a decision about the treatment is made. The dataset was obtained from the National center for biotechnology information (NCBI) and consists of 7500 influenza virus positive DNA sequences and 7500 coronavirus positive DNA sequences. The DNA sequences that this data set contains are all between 95 and 2995 characters long, with most sequences falling in the range of 1350 to 2500.

\subsection{Human immunodeficiency virus dataset}
The human immunodeficiency virus (HIV) can attack the human immune system and cause acquired immunodeficiency syndrome (AIDS). The estimated incubation period is around 8 to 9 years, during which there could be no symptoms. However, after this long period, the risk of getting opportunistic infections increases significantly and can cause many diseases. In addition to the immunosuppression, HIV can also directly have impact on the central nervous system and cause severe mental disorders \cite{perry1990organic}. There are two subtypes of HIV, namely HIV-1 and HIV-2. HIV-1 has relatively higher virulence and infectivity than HIV-2. An HIV dataset containing 1600 HIV-1 positive DNA sequences and 1600 HIV-2 positive DNA sequences, was acquired from NCBI and used to evaluate the algorithms. The sequences in this data set are between 774 and 2961 characters long.

\subsection{SARS-Cov-2 dataset}
Severe acute respiratory syndrome coronavirus 2 (SARS-Cov-2) is a subtype of coronavirus which causes coronavirus disease 2019 (Covid-19). Varying degrees of illness can be noticed among different people \cite{guan2020clinical}. The outbreak first happened in Wuhan China at the end of 2019. A few months later, the virus had spread out to many countries. As it spread very rapidly, it caused worldwide lockdowns. However, the virus seems to spread faster in the USA than in other countries, and the USA has the most confirmed cases. It is interesting to examine if the viruses found in the USA are different from those in the rest of the world. To test this, a SARS-Cov-2 dataset was obtained from NCBI. It contains 166 SARS-Cov-2 positive DNA sequences from the USA and 158 SARS-Cov-2 positive DNA sequences from the rest of the world (China, Hong Kong, Italy, France, Iran, Korea, Spain, Israel, Pakistan, Taiwan, Peru, Colombia, Japan, Vietnam, India, Brazil, Sweden, Nepal, Sri Lanka, Australia, South Africa, Greece, Turkey). The classification on this dataset was the most challenging one since the two classes are the same type of virus and there is a limited amount of data available. The DNA sequences of this data set are between 17,205 and 29,945 characters long.

\section{DNA classification methods}
In this section, the feature extraction methods and machine learning algorithms that are used are described. Two feature extraction methods were compared. The first method is based on the edit distance between two DNA strings. The second method relies on the 3-gram method \cite{yunlingdong2017}, which will be described later in detail. Six machine learning algorithms are combined with the two feature extraction methods. Finally, three state-of-the-art methods, namely a convolutional neural network (CNN), a deep neural network (DNN), and an N-gram probabilistic model, which were fed the unprocessed DNA sequences without prior feature extraction, were tested.

\subsection{Feature extraction method}

\subsubsection{Feature extraction based on edit distance}
The Levenshtein distance, also known as edit distance, is used to measure the difference between two strings. The smaller the distance, the more similar the two strings are. There are three edit operations, inserting a character into a string, deleting a character from a string, or substituting a character in a string. The Levenshtein distance between string $a$ and $b$ denotes the minimum number of edit operations that need to be performed on string $a$, in order to transform string $a$ into string $b$. The Levenshtein distance between the first $i$ characters of string $a$ and the first $j$ characters of string $b$ is denoted by $D_{ab}(i,j)$, which can be calculated using equation (1). 

\vspace*{0.4cm}

\scalebox{0.84}{
 $ \begin{array}{l}
  \begin{array}{l}D_{ab}(i,j)\;=\;\left\{\begin{array}{lc}max(i,j) \ \ ~~~~~~~~~~~~~~~ if \ \ min(i,j=0)\\min\;\left\{\begin{array}{l}D_{ab}(i-1,j)+1\\D_{ab}(i,j-1)+1 ~~~~~~~ otherwise \\D_{ab}(i-1,j-1)+1 (a_i \neq b_j)\end{array}\right.
  \end{array}\right.\\\end{array}
  \\\end{array}
  $
  }

\vspace*{0.4cm}

In this equation, $a_i$ and $b_j$ represent the $i$-th and $j$-th character in string $a$ and string $b$. If either string $a$ or string $b$ has no character, then the Levenshtein distance equals the maximum length among them. This point is easy to understand because if one string is empty, then simply inserting all characters from the other string into the empty string is enough. If both strings are not empty, then the last character in both strings, namely $a_i$ and $b_j$, should be examined. If they have the same terminal character, then both of them could be ignored. Only looking at the first $(i-1)$ characters in string $a$ and the first $(j-1)$ characters in string $b$ is enough. In this scenario, $D_{ab}(i-1, j-1)$ equals $D_{ab}(i, j)$. If the terminal characters are different, then the costs of three possible options are supposed to be compared. The first option is deleting the terminal character in string $a$. Such that $D_{ab}(i-1, j)$, which is the Levenshtein distance between the first $(i-1)$ characters in string $a$ and the first $j$ characters in string $b$ should be calculated, plus $1$ caused by the deletion. The second option is inserting one character, which should be the same as the terminal character of string $b$, to the end of string $a$. By the insertion operation, the terminal character of string $b$ can be ignored, and we only need to calculate $D_{ab}(i, j-1)+1$. The insertion causes the addition of 1. The last option is substituting the terminal character of string $a$ by the terminal character of string $b$. In this scenario, the terminal character in both strings can be ignored, and this is denoted by $D_{ab}(i-1, j-1)+1$. Formula (1) suggests a recursive procedure to calculate the Levenshtein distance between two strings. In each recursion, the last character of one or both strings could be ignored. The recursion should halt when one string is empty. According to this formula, the Levenshtein distance between two string $a$ and $b$ is denoted by $D_{ab}(|a|, |b|)$, where $|a|$ and $|b|$ represent the length of string $a$ and $b$.

As mentioned in the introduction section, primers are short DNA sequences that are used in medical research to detect possible viruses by conducting PCR on the DNA sequence. In bioinformatics, primers can be used to extract features. In order to do so, the Levenshtein distance between the primer and the DNA sequence is calculated. If the distance is small, then the similarity between the DNA sequence and the primer is considerable. In other words, a person is likely to be tested positive for the virus corresponding to the primer. However, in an infected DNA sequence, the target virogenes are small fragments hidden somewhere in the DNA sequence. Therefore, directly calculating the Levenshtein distance between the DNA sequence and the primer is not accurate. 

In this research, the matching process is done between the primer and a short sub-string of the DNA sequence. Then, the window on the DNA sequence slides by one character, and the Levenshtein distance between the primer and the next sub-string of the DNA sequence is calculated. This process is repeated until the whole DNA sequence is traversed. Finally, the minimum distance of all these calculated distances is taken and considered as the final distance. For example, the given DNA sequence is "TTTGACTCGT" and the window size is 8. The Levenshtein distance between the primer and the first sub-string "TTTGACTC" is calculated, then "TTGACTCG" and "TGACTCGT". Afterward, the minimum distance of the three distances is taken. In reality, many viruses have existed in the world for a long time, and they have mutated severely. Therefore, only calculating the Levenshtein distance between the DNA sequence and one primer is not enough to make the result accurate. However, for many viruses, multiple different primers exist. Thus, the minimum Levenshtein distance between the DNA sequence and various primers can be calculated and combined into a single feature vector. These feature vectors can then be used to train and test different machine learning methods.

Obtaining a virus's primers can be expensive and takes time, especially for a newfound virus like SARS-Cov-2 in early 2020 or viruses that have a high mutation rate. Our novel method to solve this problem is using randomly generated short DNA sequences to replace the primers. The feature extraction is then achieved by calculating the minimum Levenshtein distance between the randomly generated DNA strings and the DNA sequences needing classification. Since nothing except for the DNA strings used to calculate the minimum Levenshtein distance (primers vs. randomly generated DNA sequences) changed, the resulting feature vectors are of the same format. Different machine learning algorithms will be trained and tested using each set of feature vectors in the experiments.

In this method, before the feature vectors are fed into the machine learning algorithms, the usage of a normalization process is crucial. It is helpful to consider that the difference between smaller distances is more significant than the difference between larger distances. For example, the difference between distance 3 and 4 should be given more weight than between distance 30 and 40. Therefore, finding a suitable normalization function is necessary for this method. The elements in the feature vectors were processed with the function shown by formula (2),  where $x$ is the computed distance.
\begin{equation}
    f(x) = \frac{1}{1+x}
\end{equation}

\subsubsection{Feature extraction of 3-gram method}
Extracting the features of a DNA sequence using what will be referred to as the 3-gram method is based on knowledge from biology and medicine. Before diving into the algorithm, it is helpful to know how the human body builds the proteins that it needs. DNA sequences store the necessary information for building proteins. In biological cells, a DNA sequence is first reformed into an mRNA sequence. This process is called transcription, and mRNA works as a messenger. After this, the mRNA sequence is translated into a series of amino acids, which build up the proteins. Researchers have found that one amino acid is coded by a group of three nucleobases \cite{crick1961general}. The groups which contain three nucleobases are called codons. The corresponding translation between codons and amino acids is illustrated in Fig. 1. There are in total 64 different codons, and 61 of them can be translated into amino acids. The other three, TAA, TAG, and TGA are stop codons. They mark the halt point of translation. There are 20 possible amino acids, meaning that several different codons can be translated into the same amino acid.
\begin{figure}[!tbp]
        \includegraphics[width=7.7cm]{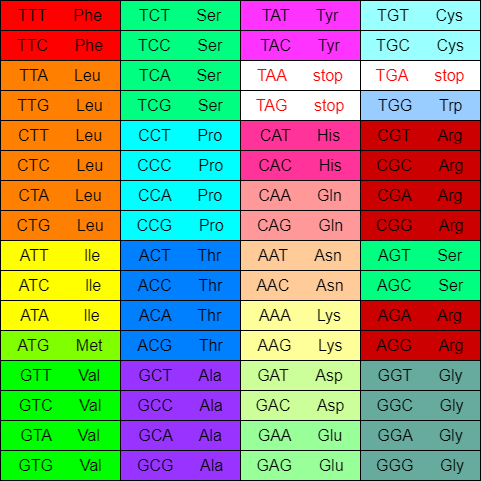}
    \caption{Various combinations of three successive nucleobases and their corresponding amino acid}
    \label{fig:afb1}
\end{figure}

The 3-gram method simulates the process from DNA sequences to amino acids. A window of size three is used to traverse the whole DNA sequence with a sliding unit of 1 at each time. At each time, the group of three nucleobases is acquired from the DNA sequence, and the corresponding amino acid is recorded. Stop codons are neglected. After the whole DNA sequence is traversed, all different types of amino acids are counted. Then the proportion of each amino acid is calculated and put in a histogram. Take the same DNA sequence as an example again. There are eight codons in the DNA sequence "TTTGACTCGT". They are "TTT", "TTG", "TGA", "GAC", "ACT", "CTC", "TCG" and "CGT". They can be translated into one phenylalanine, two leucines, one aspartic acid, one threonine, one serine, and one arginine.

In a nutshell, each DNA sequence is represented by a 20-D feature vector after using the 3-gram method. These feature vectors are then used as the input vectors for machine learning algorithms \cite{yunlingdong2017}. Unlike the previously introduced method, which extracts features based on the Levenshtein distance, the feature vectors extracted by the 3-gram method do not need to be normalized before feeding them into the machine learning algorithms. This is simply because the calculations of the proportion of each amino acid themselves already normalize the data.

\subsection{Machine learning algorithms}
The comparison between the two previously introduced feature extraction methods was done by training and testing six machine learning models on the feature vectors acquired by using the two feature extraction methods. Since all the experiments consisted of binary classification tasks, each processed vector was labeled with either 1 or 0, depending on its class. For example, whether it was associated with an infected DNA sequence or an uninfected one. The labeled data was then used to train each of the six machine learning methods described in the following subsections. K-fold cross-validation was used to ensure accurate results. The processed samples were separated into K folds randomly. The training and testing process was run K times. At each time, one fold was used as the testing set, and the remaining K-1 folds together was seen as the training set. By doing such, all folds were used as a part of the training set for K-1 times and as the testing set once. Therefore, one model was trained and tested K times, and the average test accuracy was used to evaluate the model. 

\subsubsection{Multi-layer perceptron}\label{sec:_academic_style}
The Multi-layer perceptron (MLP) is a supervised learning model which is often used in classification problems. After trying out different
architectures, we observed an MLP with three hidden layers to perform
best with the number of neurons being 500, 250, and 125, respectively. The activation function of the neurons in the hidden layers was the Rectified Linear Unit (ReLU). The sigmoid function was used as the only neuron's activation function in the output layer. 

In the training phase, the binary cross-entropy function was used as the loss function. This loss function calculates the error between the actual output and the target label, on which the training and update of the weights are based \cite{rezai2014novel}. The adaptive moment estimation optimizer (Adam) was used in this research. The learning rate of the optimization decides how fast the model learns. It is critical to the model and should be set carefully. A large learning rate might make the model never find the optimal solution, while a small learning rate causes inefficiency. Mini-batch learning was used during training. This means that several examples were fed into the MLP together, and then the weights were updated. Mini-batch learning makes sure that the learning process goes on the right track. In each epoch, all data in the training set were used. When the model was trained on all examples, the next epoch started. In preliminary experiments, it was found that the model overfitted severely. Therefore, the dropout technique was used to prevent overfitting. The key idea of dropout is that during the training phase, some units and their connections are randomly dropped, enabling the model to generalize well \cite{srivastava2014dropout}.

\subsubsection{Logistic regression}
Logistic regression is a linear model that is used to carry out binary classification. Similar to the MLP model, the sigmoid function was used for the output unit. Additionally, L2 regularization was used to prevent possible overfitting. This method adds a regularization term at the end of the loss function, which is illustrated by formula (3). The additional term is also known as L2 regularization. The hyperparameter $C$ in this term controls to what degree the L2 regularization should be executed. Smaller values specify stronger regularization. 

\begin{equation}
    Loss \leftarrow Loss + \frac{1}{C} \sum_{i}^{n}w_{i}^{2}
\end{equation}

Usually, it is assumed that the independent variables in the input vector $X$ have a multivariate normal distribution. However, most of the time, this assumption is not satisfied. In such situations, logistic regression is a good alternative model \cite{ertas2018detection}.

\subsubsection{Support vector machine}
The support vector machine (SVM) is another linear model for the classification task \cite{boser1992training,vapnik:1995}. An SVM is capable of handling a small amount of data and is less sensitive to noise in a dataset, and therefore it has an excellent generalization ability \cite{hamidi2015comparative}. The SVM aims to find the hyperplane which maximizes the margin separating the two classes. The solution to this can be found by using the Lagrange multiplier method. Powerful non-linear SVM models can be trained if kernel functions are appropriately used \cite{boser1992training}. Kernel functions create new feature vectors that usually have more dimensions than the original input. The SVM finds the new hyperplane, which is linear in the new feature space. However, in the original feature space, the separation will be non-linear if a non-linear
kernel is used. 

In the training process of an SVM, the inner product of two samples $x_i$ and $x_j$ needs to be calculated. The kernel method provides a solution that allows the model to get the inner product in the higher-dimensional feature space directly. This idea is illustrated by formula (4), where $K$ is the kernel function. $\phi(x_i)$ and $\phi(x_j)$ are the new feature vectors in the higher-dimensional feature space. By mapping like this, the transformation from the lower-dimensional space to the higher-dimensional space of each individual sample is unnecessary, which saves a lot of memory and computational resources. 
\begin{equation}
    K(x_i,x_j) = \phi(x_i)^{T}\cdot \phi(x_j) 
\end{equation}

In our implementation, the radial basis function kernel (RBF kernel) was used. The RBF kernel function is shown in formula (5). The hyperparameter $\gamma$ decides the distribution of the feature vectors in the higher-dimensional feature space. The other hyperparameter in the SVM model is $C$. Similar to the usage of $C$ in logistic regression, the hyperparameter $C$ here decides the regularization, or in other words, a penalty degree. 
\begin{equation}
    K(x_i,x_j) = e^{\frac{-||x_i-x_j||^2}{2\sigma^2}},\ \   \gamma = \frac{1}{2\sigma^2}
\end{equation}

\subsubsection{Random forest}
Before talking about the random forest algorithm, it is essential to know how a decision tree classifier \cite{quinlan1986induction} works. The decision tree is a tree-like model that simulates how humans make decisions. There is a judgment at each node, and the data are classified into different child nodes. The leaf nodes show the final results. The impurity drop is used to evaluate the decision, and a good query should maximize the impurity drop. The decision tree is also a supervised learning model in which the training set is used by the model to learn how to make queries and split the data until a specific criterion or threshold is reached.  

The decision tree is the fundamental model of the random forest algorithm \cite{breiman2001random}. As its name implies, a forest is built up from many trees. The random forest model trains multiple different decision trees on different data, and the average output of these trees is taken as the final output. The bootstrap aggregating method is used to create different training datasets. This method samples some data with replacement from the original training data set, and they are used to train one individual tree. The random forest is a simple, easy-to-understand algorithm which is capable of handling complex  non-linear classification task. Therefore, it is often used in the machine learning field. Two hyperparameters needed to be tuned in our implementation. One of them is the number of estimators. It controls how many trees should be created during the experiment. The other one is the maximum depth of each tree. This value should neither be too large nor too small. Larger numbers may cause overfitting, while lower depths could lead to underfitting.

\subsubsection{Adaboost}
Adaboost \cite{freund1995desicion}, which is short for adaptive boosting, is another decision-tree-based algorithm similar to the random forest. The basic idea of Adaboost is training multiple weak classifiers, and their combination is a more robust classifier. The similarity between random forest and Adaboost is that they both train multiple classifiers, while the difference lies in the data used to train them. Unlike random forest, which uses part of the data at each time, Adaboost uses all data in the original training set to train a single classifier. Those samples which were classified incorrectly are given more weight, and the updated data set will be used to train the next classifier. Similar to the random forest algorithm, two hyperparameters needed to be tuned in our implementation. They are the number of estimators and the maximum depth of the trees. 

\subsubsection{XGBoost}
XGBoost \cite{chen2016xgboost} is another ensemble learning algorithm like random forest and Adaboost. The difference between these three algorithms lies in how they train one individual decision tree classifier. Random forest uses different sampling data, while Adaboost manipulates the weights of data. Different from both of them, XGBoost is built based on the idea of the gradient boosting decision tree (GBDT) and developed from that. GBDT trains an individual decision tree to fit the residual from the previous decision tree \cite{chen2016xgboost}. This procedure is done by considering the whole decision tree as a function $F(x)$, and calculating the gradient of the loss function with respect to the function $F(x)$. XGBoost introduces a regularization term to the loss function to prevent overfitting. Additionally, each leaf node is given a score, such that the loss function can be computed more efficiently. XGBoost enables researchers to solve large-scale problems in the real world by using a relatively small amount of resources \cite{chen2016xgboost}. In our implementation, three hyperparameters needed to be tuned. They are the number of trees that should be trained, the maximum depth in each tree, and the $\lambda$ value which controls the degree of regularization.

\subsection{State-of-the-art algorithms}
In this section, three more complex and computationally expensive state-of-the-art algorithms are described that will be compared to the simpler machine learning algorithms using the two feature extraction methods. For these more complex algorithms, the DNA sequences are used as feature vectors without the in-between step of feature extraction via one of the previously described methods.

\subsubsection{Convolutional neural network}
With the development of computational capacity, the convolutional neural network (CNN) \cite{lecun1989generalization,lecun1998gradient} has been widely used in many fields such as computer vision and has achieved great success. A CNN is a neural network in which some matrix multiplication operations between layers are replaced by convolutions \cite{goodfellow2016deep}. The CNN is able to learn to extract features and train a classifier at the same time.  Recently, several researchers have applied the CNN model to bioinformatics, especially the task of classifying DNA sequences. The CNN model has been found to be capable of handling the classification of the nucleotides of DNA sequences with A, G, C, and T \cite{nguyen2016dna}. Another study proved the feasibility of using CNNs to classify non-coding RNA sequences, and accuracies higher than 95\% were achieved on multiple datasets \cite{aoki2018convolutional}. Based on these previous studies,  we developed a CNN model and compared it to the other algorithms.

Before training a CNN model, an individual DNA sequence should be transformed into a 2-dimensional matrix by using one-hot encoding. In the resulting matrix, each column represents a character in the original DNA sequence. The number 1 appears at the place which stands for the corresponding character, while the other places in the same column are filled with 0. An example of one-hot encoding is illustrated in Fig. 2. The first to the last position in the column represents A, C, G, T, respectively. By using the one-hot encoding, each character in the original DNA sequence is represented by 4 channels. The channels are shown below each other in the same column in Fig. 2. Since the DNA sequences have different lengths, all of them are padded with columns with only zeros, to the same length. 
\begin{figure}[!htb]
    \includegraphics[width=7.7cm]{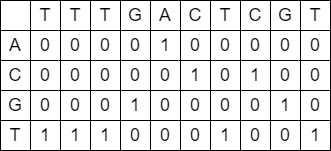}
    \caption{Example of one-hot encoding of the DNA sequence "TTTGACTCGT"}
    \label{fig:afb1}
\end{figure}

In the convolution layer, a neuron uses a kernel (filter) and performs a convolution operation to compute a single output in the resulting feature map. Afterward, the filter slides to the next region and repeats the convolution operation. In this way, the features from different parts of the input can be extracted. In order to decrease the size of the feature maps, a max-pooling layer is added after the convolution layer. It reduces the size by only keeping the maximum value from several neighboring values in a feature map. 

After experimeting with different CNN architectures, finally nine convolution layers were used for the SARS-Cov-2 dataset and seven were used for the others. Each convolution layer was followed by a max-pooling layer. One hundred filters were used in each convolution layer. Each filter in all layers received all four channels as input and has a window length of 3. Therefore, a filter did not only take one character each time but integrated them together with a width of 3. The stride in all convolution layers was set to 1, while in all pooling layers the stride was set to 3. All the outputs resulting from the last max-pooling layer were fully connected to a dense layer with five hundred neurons. The final output layer with a single output followed the dense layer. Similar to the MLP model that we implemented, ReLU was used as the activation function on the neurons in the convolution layer and dense layer. The activation function used in the output layer was the sigmoid function. The loss function of this model was the binary cross-entropy function. 

Similar to the MLP, three hyperparameters were coarsely tuned. They were the number of epochs, the batch size, and the learning rate of the gradient descent optimization. These hyperparameters also apply to the DNN algorithm that will be described next. 

\subsubsection{Deep neural network}
Another approach is to use a multi-layer perceptron with multiple hidden layers, also referred to as a deep neural network or DNN. DNNs have been increasingly used successfully in bioinformatics. An earlier study demonstrated that a DNN could perform with a state-of-the-art level at the task of predicting species-of-origin and species classification of short DNA sequences \cite{busia2019deep}. The training phase of DNNs uses each of the DNA sequences in their entirety, and also does not use the features extracted by either taking minimum Levenshtein distances or the 3-gram method. Similar to the CNNs, the DNA sequences were transformed into a 2-dimensional matrix using one-hot encoding. After testing multiple different configurations, a model consisting of one dense layer with 40 neurons, followed by another dense layer with 20 neurons appeared to provide the most accurate results. For this model, the activation function of the neurons in the hidden layers was again ReLU, and the one for the output layer was the sigmoid function. The loss function used was the binary cross-entropy function as well.

\subsubsection{N-gram probabilistic model}
A third state-of-the-art method that uses the entire DNA sequences as the features it is trained on is an N-gram probabilistic model. This method is commonly used in natural language processing (NLP) \cite{tremblay2011effects}. It has also been successfully applied to DNA classification problems \cite{tomovic2006n,chen2014private} with resulting classification accuracies up to 99.6\%. An N-gram is a sequence of N items. These items are e.g. words in NLP or the letters A, C, G, or T, representing nucleobases in DNA sequences in FASTA format. The N-gram probabilistic model can be used to predict the probability of the next item $x$ in a sequence given the history of the n-1 previous items $h$: $P(x|h)$. It can for example be used to predict the probability of the next item being the letter A in a DNA sequence, given that the previous four letters were "ACGT". This probability would be calculated as $P(A|G, T)$ in an N-gram model with an N-value of 3 and $P(A|A, C, G, T)$ for one with an N-value of 5. 

For our experiments, the N-gram probabilistic model was used as a classifier, so the probabilities of the next item were calculated using the previous N-1 items for each class separately, e.g. $P(X_t|X_
{t-1}, X_{t-2}, Class)$ for N=3. Eventually, the prior probability $P(Class)$ can be used in combination with the N-gram probabilistic model(s) to compute the probability of a sequence belonging to a certain class using Bayes's rule:

\vspace*{0.4cm}
{\small
$P(Class|X_1, X_2, ...X_T) =  P(Class) * P(X_2| X_1, Class)\\ 
                    * P(X_3 | X_2, X_1, Class)...
                 * P(X_t | X_{t-1}, X_{t-2}, Class)...
                     * P(X_T | X_{T-1}, X_{T-2}, Class)$

}
\vspace*{0.4cm}

In order to prevent underflow when working with DNA sequences containing more than 30,000 items, in this experiment, the probabilities were not multiplied, but the logarithmic values of these probabilities were added.


The number of occurrences of each N-gram is counted for each class. These counts are then used to calculate the probabilities.
For testing purposes on a novel DNA sequence, the computed probabilities are  used to calculate the log probability of belonging to both classes. The DNA sequence is then classified according to the class with the
highest probability. We tested different values for N, and the final 
best performing value of N that was used in all experiments was 6.

\section{Experimental setup}
Each machine learning algorithm was trained and tested on the processed feature vectors obtained by using the two different feature extraction methods. Reliable primers of HCV could be acquired. Therefore, when testing the random DNA sequences method on the HCV dataset, the feature extraction using primers based on distance was done in order to make comparisons with the random DNA sequences method. Since 37 primers of HCV were acquired, we generated three groups of random DNA sequences, and each contains 37 DNA sequences. The lengths of the DNA sequences of the three groups were 25, 100, and 200, respectively. Additionally, a fourth group was generated, in which there were 100 DNA sequences of length 200. For other datasets, since the primers could not be obtained, the random DNA sequences method was tested using 50 random DNA sequences, with a length of 25, 50, and 100. 

Furthermore, each algorithm was trained and tested on each of the four datasets using the DNA sequences in their entirety as features. For every experiment, the accuracy of the binary classification was tested across ten folds of cross-validation. As all datasets consisted of two kinds of DNA sequences, the training and testing procedure was the same for each dataset. 

For each experiment, the feature vectors were assigned labels according to their class. During the training phase, the classifiers were trained on each feature vector of the training set and its corresponding label. During the testing phase, each feature vector of the testing set was classified as either 'positive' or 'negative' for the HCV dataset, HIV-1 or HIV-2 for the HIV dataset 'influenza virus' or 'coronavirus' for the influenza/corona dataset, or as either 'originating from the USA' or 'not originating from the USA' for the SARS-Cov-2 dataset. The result of each classification was recorded and compared to the correct label of each feature vector in the testing set to calculate the accuracy. For each trial, the accuracy was recorded to compute the mean accuracy and standard deviation across the ten folds of the cross-validation.

The whole experiment was divided into two parts. The first part was a preliminary experiment, which was used to tune the hyperparameters and decide the best set of hyperparameters of each algorithm. This process is done by repeating the training and testing procedure using different sets of hyperparameters. The one which gave the highest accuracy was selected. After the optimal hyperparameters were found, the second experiment was conducted using those found hyperparameters. The comparison across different algorithms was based on the results of the second experiment. All the hyperparameters that needed to be tuned have already been discussed in the method section, and the following Tables (1) to (7) show the best found values for the hyperparameters.

\begin{table*}[!htb]
\begin{tabular}{ccccccc}
\hline
 & \multicolumn{3}{c}{Random DNA method} & \multicolumn{3}{c}{3-gram method} \\
 &  Batch size     &    Epoch number   & learning rate     & Batch size      &  Epoch number     &   Learning rate      \\ \hline
HCV &   64    &  2000     & 0.0001     &    64   &  1000     & 0.0001     \\
HIV &   256    &    2000   &   0.0001   &   256    &    2000   &    0.0001  \\
Inf./Cor. &     250  &  2000     &  0.0001    & 64      &  50     & 0.0001     \\
SARS-Cov-2 &    64   &  10000     & 0.0001     &    64   &  20000     & 0.00005     \\ \hline
\end{tabular}
\caption{The best hyperparameters of the MLP on the four datasets, using the two feature extraction methods}
\end{table*}

\begin{table*}[!htb]
\begin{tabular}{ccccc}
\hline
 & \multicolumn{2}{c}{Random DNA method} & \multicolumn{2}{c}{3-gram method} \\
 &     Distribution $\gamma$      &     Regularization C     &      Distribution $\gamma$      &     Regularization C     \\ \hline
HCV &   750        &    7      &     750      &     7     \\
HIV &   200        &    10      &    200        &    10      \\
Inf./Cov. &     200      &      10    &     200      &      10    \\
SARS-Cov-2 &     1500000      &     10     &    2500000       &     10     \\ \hline
\end{tabular}\caption{The best hyperparameters of the SVM on the four datasets, using the two feature extraction methods}
\end{table*}

\begin{table*}[!htb]
\begin{tabular}{ccc}
\hline
 & Random DNA method & 3-gram method \\
 & Regularization C & Regularization C \\ \hline
HCV & 5000000 & 1000 \\
Inf./Cor. & 100000 & 100000 \\
HIV & 100000 & 100000 \\
SARS-cov-2 & 10000000 & 10000000 \\ \hline
\end{tabular}
\caption{The best hyperparameters of logistic regression on the four datasets, using the two feature extraction methods}
\end{table*}

\begin{table*}[!htb]
\begin{tabular}{ccccc}
\hline
 & \multicolumn{2}{c}{Random DNA method} & \multicolumn{2}{c}{3-gram method} \\
 &     Number of trees      &     Max depth     &      Number of trees      &     Max depth     \\ \hline
HCV &   50        &    50      &     50      &     50     \\
HIV &   50        &    50      &    50        &    50      \\
Inf./Cov. &     25      &      50    &     25      &      50    \\
SARS-Cov-2 &     100      &     50     &    100       &     50     \\ \hline
\end{tabular}\caption{The best hyperparameters of random forest on the four datasets, using the two feature extraction methods}
\end{table*}

\begin{table*}[!htb]
\begin{tabular}{ccccc}
\hline
 & \multicolumn{2}{c}{Random DNA method} & \multicolumn{2}{c}{3-gram method} \\
 &     Number of trees      &     Max depth     &      Number of trees      &     Max depth     \\ \hline
HCV &   250        &    3      &     250      &     3     \\
HIV &   250        &    3      &    250        &    3      \\
Inf./Cov. &     150      &      3    &     150      &      3    \\
SARS-Cov-2 &     250      &     3     &    250       &     3     \\ \hline
\end{tabular}\caption{The best hyperparameter of Adaboost on the four datasets, using the two feature extraction methods}
\end{table*}

\begin{table*}[!htb]
\begin{tabular}{ccccccc}
\hline
 & \multicolumn{3}{c}{Random DNA method} & \multicolumn{3}{c}{3-gram method} \\
 &  Number of trees    &    Max depth   &  $\lambda$     & Number of trees     &    Max depth   &  $\lambda$     \\ \hline
HCV &   200    &  3     & 0.25     &    200   &  3     & 0.25     \\
HIV &   200    &    3   &   0.25   &   200    &    3   &    0.25  \\
Inf./Cor. &     100  &  3     &  0.5    & 100      &  3     & 0.5     \\
SARS-Cov-2 &    300   &  3     & 0.5     &    300   &  3     & 0.25     \\ \hline
\end{tabular}
\caption{The best hyperparameters of XGBoost on the four datasets, using the two feature extraction methods}
\end{table*}

\begin{table*}[!htb]
\begin{tabular}{ccccccc}
\hline
 & \multicolumn{3}{c}{CNN} & \multicolumn{3}{c}{DNN} \\
 &  Batch size     &    Epoch number   & Learning rate     & Batch size      &  Epoch number     &   Learning rate      \\ \hline
HCV &   100    &  200     & 0.0001     &    100   &  170     & 0.00005     \\
HIV &   128    &    100   &   0.0001   &   100    &    170   &    0.0001  \\
Inf./Cor. &     32  &  25     &  0.0001    & 100      &  150     & 0.00005     \\
SARS-Cov-2 &    32   &  500     & 0.00001     &    150   &  100     & 0.0001     \\ \hline
\end{tabular}
\caption{The best hyperparameters of CNN and DNN on the four datasets}
\end{table*}

\section{Results}

\begin{figure*}[!h]
    \includegraphics[width=16cm, height=7.5cm]{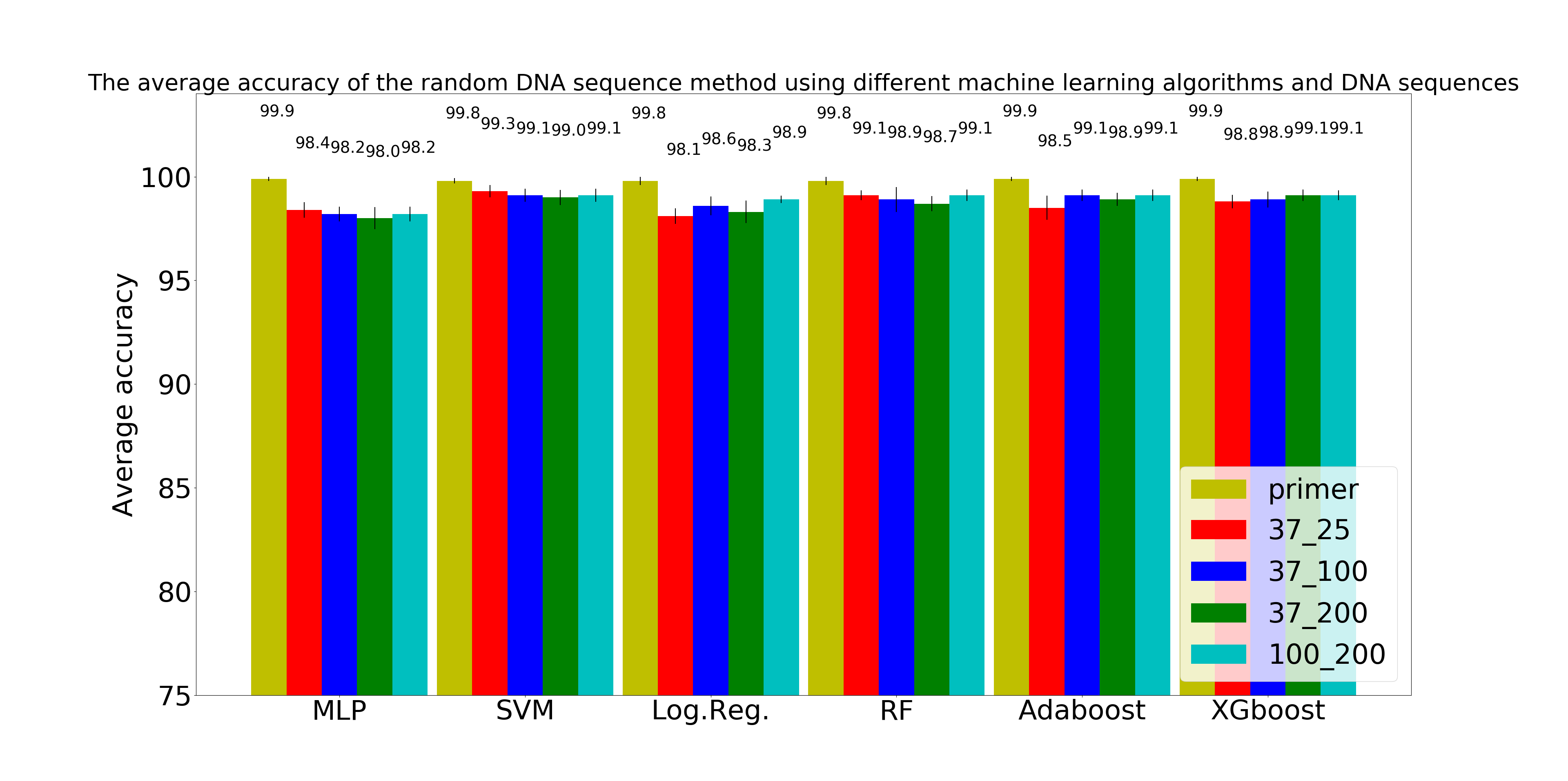}
    \caption{Accuracy of different machine learning algorithms using the primers and the novel random DNA-sequence feature extraction method on the HCV dataset}
    \label{fig:randomHCV}
\end{figure*}

All results presented in this section are the mean accuracy and standard deviation over ten folds of cross-validation. The comparison of using primers and random DNA sequences was only made on the HCV dataset. The results of the experiment on the HCV dataset are shown in Figure (\ref{fig:randomHCV}).

For each data set, the results of all six machine learning algorithms using the random DNA sequence feature extraction method are presented in Table (\ref{table:random}) containing mean accuracy and standard deviation over the ten folds of the cross-validation. For each machine learning algorithm, multiple lengths and amounts of the random DNA sequences were considered. However, only the ones showing the best results are displayed in the table. For the displayed results of the random sequence feature extraction method, 50 randomly generated DNA sequences of length 25 were used on the HIV and influenza/corona data sets. For the SARS-Cov-2 data set, 50 randomly generated DNA sequences of length 50 were used. For the 3-gram feature extraction method, the results of the six machine learning algorithms for each of the four data sets are displayed in Table (\ref{table:3gram}). 

\begin{table*}[!h]
\centering
\begin{tabular}{ccccc}
\hline
                        & HCV        & HIV        & Inf./Cor.  & SARS-Cov-2 \\ \hline
\multicolumn{1}{c}{MLP}       & $97.60 \pm 0.42$ & $99.00\pm 1.08$ & $99.73\pm 0.18$ & $91.48\pm 4.14$ \\ 
\multicolumn{1}{c}{SVM}       & $\mathbf{99.30 \pm 0.95}$ & $99.72\pm 0.37$ & $\mathbf{99.77\pm 0.16}$ & $\mathbf{97.26\pm 3.34}$ \\ 
\multicolumn{1}{c}{Log.-Reg.} & $98.90 \pm 0.57$ & $98.19\pm 0.99$ & $96.15\pm 0.38$ & $94.76\pm 2.88$ \\ 
\multicolumn{1}{c}{RF}        & $99.10 \pm 0.74$ & $99.63\pm 0.25$ & $99.40\pm 0.39$ & $95.09\pm 3.27$ \\ 
\multicolumn{1}{c}{Adaboost}  & $99.10 \pm 0.88$ & $\mathbf{99.88\pm 0.16}$ & $99.62\pm 0.13$ & $95.37\pm 5.08$ \\ 
\multicolumn{1}{c}{XGBoost}   & $99.10 \pm 0.74$ & $99.84\pm 0.22$ & $99.66\pm 0.20$ & $94.14\pm 1.71$ \\ \hline
\end{tabular}
\caption{Mean accuracy $\pm$ standard deviation for all methods using the random DNA-sequence feature extraction across the four data sets. }
\label{table:random}
\end{table*}

\begin{table*}[!h]
\centering
\begin{tabular}{ccccc}
\hline
                          & HCV        & HIV        & Inf./Cor.  & SARS-Cov-2 \\ \hline
\multicolumn{1}{c}{MLP}       & $99.70\pm 0.48$ & $99.84\pm 0.22$ & $\mathbf{99.99\pm 0.02}$ & $86.82\pm 6.56$ \\ 
\multicolumn{1}{c}{SVM}       & $\mathbf{99.90\pm 0.32}$ & $99.88\pm 0.26$ & $\mathbf{99.99\pm 0.02}$ & $\mathbf{92.90\pm 3.82}$ \\ 
\multicolumn{1}{c}{Log.-Reg.} & $99.40\pm 0.84$ & $99.60\pm 0.36$ & $\mathbf{99.99\pm 0.02}$ & $87.97\pm 6.53$ \\ 
\multicolumn{1}{c}{RF}        & $99.70\pm 0.84$ & $99.75\pm 0.20$ & $99.98\pm 0.03$ & $91.04\pm 4.72$ \\ 
\multicolumn{1}{c}{Adaboost}  & $99.80\pm 0.42$ & $\mathbf{99.91\pm 0.15}$ & $\mathbf{99.99\pm 0.02}$ & $92.87\pm 4.27$ \\ 
\multicolumn{1}{c}{XGBoost}   & $\mathbf{99.90\pm 0.32}$ & $99.66\pm 0.34$ & $99.97\pm 0.03$ & $89.86\pm 6.84$ \\ \hline
\end{tabular}
\caption{Mean accuracy $\pm$ standard deviation for all methods using the 3-gram feature extraction across the four data sets. }
\label{table:3gram}
\end{table*}

An overview of all results, in which the feature extraction methods are compared with state-of-the-art algorithms, are provided in Table (\ref{table:all}). For this table, only the best results of the 3-gram and random DNA sequence feature extraction method were considered. The best results stem from different machine learning algorithms for different data sets. However, the exact results for each algorithm on each data set are displayed in Table (\ref{table:random}) and Table (\ref{table:3gram}). 

Figure (\ref{fig:randomHCV}) suggests that using primers has the highest accuracy when the classifier is trained by using the MLP, Adaboost, or XGBoost algorithm. If the primers are replaced by random DNA sequences, the highest accuracy is obtained using an SVM classifier. Although
using primers leads to better results, the results indicate that using primers (M=99.9, SD=0.32) does not have a significantly higher accuracy (t(18)=1.90, p=0.07) than using the random DNA sequences (M=99.3, SD=0.95). It can be concluded that the Levenshtein distance feature extraction yields the best and most consistent results across the six different machine learning algorithms when the distance between a primer and a DNA sequence is taken. However, the random DNA sequences can be used to replace primers when they are not available. 

Furthermore, it can be observed that even though the SVM produces the highest accuracy for three of the four data sets, there is not one machine learning method that consistently yields the best results across all the different lengths of the randomly generated strings nor across each of the various data sets (see Table (\ref{table:random})). Also, there is no clear indication about the best length of the random DNA strings
(for simplicity, we do not show all these results).

\begin{table*}[!htb]
\centering
\begin{tabular}{ccccc}
\hline
                             & HCV                 & HIV                 & Inf./Cor.           & SARS-Cov-2          \\ \hline
\multicolumn{1}{c}{Random} & $99.30\pm 0.95$          & $99.88\pm 0.16$          & $99.77\pm 0.16$          & $\mathbf{97.26\pm 3.34}$          \\ 
\multicolumn{1}{c}{3-gram} & $\mathbf{99.90\pm 0.32}$          & $\mathbf{99.91\pm 0.15}$ & $\mathbf{99.99\pm 0.02}$ & $92.90\pm 3.82$          \\ 
\multicolumn{1}{c}{CNN}    & $99.20\pm 1.03$          & $99.81\pm 0.16$          & $99.97\pm 0.03$          & $88.82\pm 5.13$          \\ 
\multicolumn{1}{c}{DNN}    & $98.50\pm 1.08$          & $99.47\pm 0.42$          & $99.79\pm 0.11$          & $90.12\pm 4.11$          \\ 
\multicolumn{1}{c}{N-gram} & $99.10\pm 1.29$          & $99.78\pm 0.26$          & $99.97\pm 0.05$         & $90.15\pm 5.37$       \\ \hline
\end{tabular}%
\caption{Mean accuracy $\pm$ standard deviation for all methods across the four data sets. }
\label{table:all}
\end{table*}

For the 3-gram feature extraction method, the results show a similar pattern. Even though here the SVM is among the machine learning algorithms yielding the best results for three out of the four data sets, Adaboost provides the highest mean accuracy for the HIV data set again. A significant difference to the random DNA string feature extraction is that the difference between the machine learning algorithms becomes much smaller. Different algorithms show identical results for two of the four data sets using the 3-gram feature extraction method.

Table \ref{table:all} shows that overall the 3-gram feature extraction method combined with either an SVM (for HCV and Inf./Cor.) or Adaboost (for HIV) obtains the highest mean accuracy of all tested methods in 3 out of 4 data sets. For these datasets also the other methods perform
very well, especially the CNN, and the differences are quite small.
The DNN seems to perform a bit worse on most datasets.

For the SARS-Cov-2 data set, the Levenshtein distance with
the random DNA string feature extraction method obtains significantly
higher accuracies than the other methods. For this small dataset, 
it outperforms the second best method (the 3-gram) with around 4.3\%.

The above results suggest that the 3-gram method obtains better performances on larger datasets, while the random DNA sequences method might be better at handling relatively smaller datasets. If large amounts of data are not readily available, the results of the random DNA sequence method are promising. It obtains an accuracy as high as 97\% 
with as little as 292 samples to train on.

\section{Conclusion and Discussion}\label{sec:conclusions}
This paper aimed to provide an extensive comparison of different methods for DNA sequence classification. Five different methods were compared across four different data sets of various sizes. Examining the proposed novel method using random DNA sequences to extract features based on distance is one of the main novelties in this paper. We wanted to test whether it is good enough to replace primers.

The results showed that modern state-of-the-art methods from fields like computer vision and natural language processing as CNNs or N-gram probabilistic models can achieve very high accuracies above 99\% on DNA sequence classification problems provided that enough sample data is available. Although the DNN has a slightly worse performances in some of the experiments, the achievements are acceptable. Therefore, we can conclude that these algorithms can be successfully applied to different
DNA classification problems. 

The results also showed that the use of feature extraction methods 
is useful to obtain the best results. The 3-gram method is quite simple
but very effective in handling different datasets. The novel feature
extraction method based on random DNA sequences led to the best
result on the smallest SARS-Cov-2 dataset and can therefore be
promising for DNA classification problems when little data is 
available.

The potential applications of the proposed methods are plenty. A potential field in which the methods could be deployed is the diagnosis of diseases. Especially the 3-gram feature extraction method seems promising to be used for diagnosing viral infections such as HCV or HIV. For future studies, it would be interesting to investigate some further applications of different methods. For example, the field of ancestral research using genetic samples or the detection of genetic predispositions are possible applications. If the same techniques perform similarly well for problems of this kind needs to be determined. Our results also indicate that the SARS-Cov-2 viruses spreading in the USA seems to be different from other countries. Therefore, it would be interesting for biologists to further investigate the origin of SARS-Cov-2 with the help of machine learning.

\bibliographystyle{plain}


%
%



\end{document}